\documentclass{jnmp}

\newcommand{\pard}[2]{\frac{\partial #1}{\partial #2}}

\def\CRAS{C.~R.~Acad.~Sc.~Paris}
\def\PTP{Prog.~Theor.~Phys.~}

\def \D {\hbox{d}}
\def\cKdV{c-KdV}

\def\Kaone {K_{1,{\rm a}}}

\def\QA{\tilde Q_1}
\def\QB{\tilde Q_2}
\def\PA{\tilde P_1}
\def\PB{\tilde P_2}
\def\Qj{\tilde Q_j}
\def\Pj{\tilde P_j}

\JNMPnumberwithin{equation}{section}

\theoremstyle{definition}

\begin{document}

%

\renewcommand{\evenhead}
 {Robert Conte, Micheline Musette, and Caroline Verhoeven}
\renewcommand{\oddhead}
 {Explicit integration of the H\'enon-Heiles Hamiltonians}

%
\thispagestyle{empty}

\FirstPageHead{200*}{*}{*}{\pageref{firstpage}--\pageref{lastpage}}{Article}

\copyrightnote{200*}
 {Robert Conte, Micheline Musette, and Caroline Verhoeven}

\Name
 {Explicit integration of the H\'enon-Heiles Hamiltonians
 \footnote{
 Corresponding author RC.
 S2004/044.
 Special issue in honour to Francesco Calogero.
 }
 }

\label{firstpage}

\Author
{
Robert CONTE~$^\dag$,
Micheline MUSETTE~$^\ddag$
and
Caroline VERHOEVEN~$^\ddag$
}

\Address{
$^\dag$ Service de physique de l'\'etat condens\'e
(Unit\'e de recherche associ\'ee au CNRS no.~2464)
\\~~CEA--Saclay, F--91191 Gif-sur-Yvette Cedex, France
\\~~E-mail: Conte@drecam.saclay.cea.fr\\[10pt]
$^\ddag$ Dienst Theoretische Natuurkunde, Vrije Universiteit Brussel
and
International Solvay Institutes for Physics and Chemistry
\\~~Pleinlaan 2, B--1050 Brussels, Belgium
\\~~E-mail: MMusette@vub.ac.be, CVerhoev@vub.ac.be
}

\Date{21~December~2004}

\begin{abstract}
\noindent
We consider the cubic and quartic H\'enon-Heiles Hamiltonians
with additional inverse square terms,
which pass the Painlev\'e test for only seven sets of coefficients.
For all the not yet integrated cases
we prove the singlevaluedness of the general solution.
The seven Hamiltonians enjoy two properties:
meromorphy of the general solution,
which is hyperelliptic with genus two
and
completeness in the Painlev\'e sense
(impossibility to add any term to the Hamiltonian without destroying
the Painlev\'e property).
\end{abstract}


\section{Introduction}
\indent

The ``H\'enon-Heiles Hamiltonian'' (HH) \cite{HH}
originally denoted a two-degree of freedom classical
Hamiltonian, the sum of a kinetic energy and a potential energy,
in which the potential is a cubic polynomial in the position variables
$q_1,q_2$,
\begin{eqnarray}
H
& =&
 \frac{1}{2} (p_1^2 + p_2^2+ q_1^2 + q_2^2)+ q_1 q_2^2 - \frac{1}{3} q_1^3,\
\label{eqHHOriginal}
\end{eqnarray}
i.e.~the simplest one after two coupled harmonic oscillators.
This system, which describes the motion of a star in the axisymmetric potential
of the galaxy ($q_1$ is the radius, $q_2$ is the altitude),
happens to be nonintegrable and to display a strange attractor.
However, if one changes the numerical coefficients,
the system may become integrable in some sense,
and this question (to find all the integrable cases and to integrate them)
has attracted a lot of activity in the last thirty years.

The present article is a self-contained paper which reviews
the current state of this problem, restricted here to the autonomous case.
It covers some old results for completeness
and it presents an explicit integration for all the cases which
have not yet been integrated.
To summarize
the possibly integrable cases (in the Liouville sense or in the Painlev\'e
sense) have been isolated long ago
and the explicit integration of all these cases
is now achieved in the Painlev\'e sense
(finding a closed-form single-valued expression for the general solution)
but not yet in the Hamilton-Jacobi sense
(finding the separating variables of the Hamilton-Jacobi equation).

In section \ref{sectionIntegrability}
we briefly recall the three main accepted meanings
of the word \textit{integrability} for Hamiltonian systems.
In section \ref{sectionOne_degree},
taking one degree of freedom as an example,
we relax the requirement on $q_j$
by accepting the singlevaluedness of some integer power of $q_j$.
In section \ref{sectionThe_seven}
one recalls the results of the Painlev\'e test for two degrees of freedom,
i.e.~the selection of three ``cubic'' cases
plus four ``quartic'' cases.
In section \ref{sectionLink}
we recall the link of these seven cases with soliton systems
and we present in the quartic case a fourth-order first-degree
ordinary differential equation (ODE)
equivalent to the Hamilton's equations.
In section \ref{sectionStrategies}
we enumerate three possible strategies to prove the singlevaluedness of the
general solution.
In section \ref{sectionIntegrationStaeckel}
we briefly recall the separation of variables in the two St\"ackel cases
(one cubic and one quartic).
In section \ref{sectionIntegrationCubicSKandKK}
the two remaining cubic cases are integrated by separating the
Hamilton-Jacobi equation.
In the three remaining quartic cases,
sections \ref{sectionIntegrationQuartic1:6:1and1:6:8}
and \ref{sectionIntegrationQuartic1:12:16},
the singlevaluedness is proven by building a birational transformation
to ODEs in the classification of Cosgrove \cite{Cos2000a}.

\section     {Integrability of Hamiltonian systems}
\label{sectionIntegrability}
\indent

Given a Hamiltonian system with a finite number $N$ of degrees of freedom,
three main definitions of \textit{integrability} are known for it,
\begin{enumerate}
\item
the one in the sense of Liouville,
that is the existence of $N$ independent invariants $K_j$
the pairwise Poisson brackets of which vanish,
$\left\lbrace K_j,K_l \right\rbrace=0$,
\item
the one in the sense of Hamilton-Jacobi,
\index{separating variables}
\index{Arnol'd-Liouville}
which is to find explicitly some canonical variables $s_j,r_j,j=1,N$
which ``separate'' the
\textit{Hamilton-Jacobi equation} for the action $S$
\cite[chap.~10]{ArnoldMechanics},
 \index{Hamilton-Jacobi equation}
which for two degrees of freedom is
\begin{eqnarray}
& &
H(q_1,q_2,p_1,p_2)-E=0,\
p_1=\pard{S}{q_1},\
p_2=\pard{S}{q_2},
\label{eqHJAutonomous}
\end{eqnarray}
\item
the one in the sense of Painlev\'e \cite{Cargese1996Conte},
i.e.~the proof that the general solution $q_j(t)$
is a single-valued expression of the time $t$,
represented either by an explicit, closed-form expression,
or by the solution of a Jacobi inversion problem,
see e.g.~(\ref{eqHyperellipticGenusTwoSystem}) below.
In the particular case $N=2$,
the inversion of the system of two integrals
\begin{eqnarray}
{\hskip -15.0truemm}
& &
C_1 =  \int_{\infty}^{s_1} \frac{\D s}{\sqrt{P(s)}}
      +\int_{\infty}^{s_2} \frac{\D s}{\sqrt{P(s)}},
\qquad
t+C_2= \int_{\infty}^{s_1} \frac{s \D s}{\sqrt{P(s)}}
      +\int_{\infty}^{s_2} \frac{s \D s}{\sqrt{P(s)}},
\end{eqnarray}
in which $P$ is a polynomial of degree $5$ or $6$,
and $C_1, C_2$ are two constants of integration,
leads to symmetric functions of $s_1, s_2$ being
meromorphic in $t$.

\end{enumerate}

\textit{Remark}.
One can prove \cite{BV1990}
that every Liouville integrable system has a Lax pair,
and the Lax pair is indeed the starting point of a powerful method
\cite{Sklyanin1995} to compute the separating variables.

The goal here is to integrate in the sense of Painlev\'e,
and ideally to find the separating variables of the Hamilton-Jacobi
equation.

\section     {The case of one degree of freedom}
\label{sectionOne_degree}
\indent
In this case,
$H=p^2/2 + V(q)$,
the Hamilton's equations of motion
admit a singlevalued general solution if and only if
$V$ is a polynomial of degree at most four,
in which case $q(t)$ is an elliptic function or one of its degeneracies.
If one slightly relaxes the requirement on $q$
and accepts that only some integer power $q^n$ be singlevalued,
then one additional case arises \cite{GambierThese},
\begin{description}
\item
$n=\pm 2$ and $V$ the sum of even terms \cite{Ermakov,Pinney},
\begin{eqnarray}
& &
H=\frac{p^2}{2} + a q^2 + b q^4 + c q^{-2},
\end{eqnarray}
in which case $q^2$ obeys 
either a linear equation ($b=0$) 
or the Weierstrass elliptic equation ($b\not=0$).
\end{description}

\section     {Two degrees of freedom: the seven H\'enon-Heiles Hamiltonians}
\label{sectionThe_seven}
\indent
If one considers the most general
two-degree of freedom classical autonomous Hamiltonian
\begin{eqnarray}
H
& =&
 \frac{1}{2} (p_1^2 + p_2^2) + V(q_1,q_2)
\label{eqH2TV}
\end{eqnarray}
and if one requires the existence of some singlevalued
integer powers $q_1^{n_1}, q_2^{n_2}$,
it is then necessary that the Hamilton's equations of motion,
when written in these variables,
pass the Painlev\'e test.
The application of this test isolates two classes of potentials $V$,
called ``cubic'' and ``quartic'' for simplification.
\begin{enumerate}
\item
In the cubic case HH3 the admissible Hamiltonians are,
\cite{CTW,Fordy1991,CFP1993},
       \index{H\'enon-Heiles Hamiltonian!cubic}
\begin{eqnarray}
H
& =&
 \frac{1}{2} (p_1^2 + p_2^2 + \omega_1 q_1^2 + \omega_2 q_2^2)
    + \alpha q_1 q_2^2 - \frac{1}{3} \beta q_1^3
    + \frac{1}{2} \gamma q_2^{-2},\
\alpha \not=0
\label{eqHH0}
\\
& & q_1'' + \omega_1 q_1 - \beta q_1^2 + \alpha q_2^2 = 0,
\label{eqHH1}
\\
& & q_2'' + \omega_2 q_2 + 2 \alpha q_1 q_2 - \gamma q_2^{-3}
=0,
\label{eqHH2}
\end{eqnarray}
in which the constants $\alpha,\beta,\omega_1,\omega_2$ and $\gamma$
can only take the three sets of values,
\begin{eqnarray}
\hbox{(SK)} : & & \beta/ \alpha=-1, \omega_1=\omega_2,\
\label{eqHH3SKcond} \\
\hbox{(KdV5)} : & & \beta/ \alpha=-6,\
\label{eqHH3K5cond}\\
\hbox{(KK)} : & & \beta/ \alpha=-16, \omega_1=16 \omega_2.
\label{eqHH3KKcond}
\end{eqnarray}
The meaning of the labels SK, KdV5, KK are explained in section
\ref{sectionLink}.
\item
In the quartic case HH4
the admissible Hamiltonians are,
\cite{RDG1982,GDR1983},
\index{H\'enon-Heiles Hamiltonian!quartic}
\begin{eqnarray}
H & = &
\frac{1}{2}(P_1^2+P_2^2+\Omega_1 Q_1^2+\Omega_2 Q_2^2)
 +C Q_1^4+ B Q_1^2 Q_2^2 + A Q_2^4
\nonumber
\\
& &
 +\frac{1}{2}\left(\frac{\alpha}{Q_1^2}+\frac{\beta}{Q_2^2}\right)
 + \gamma Q_1,\ B \not=0,
\label{eqHH40}
\\
& & Q_1''+\Omega_1 Q_1 + 4 C Q_1^3 + 2 B Q_1 Q_2^2 - \alpha Q_1^{-3} + \gamma=0,
\label{eqHH41}
\\
& & Q_2''+\Omega_2 Q_2 + 4 A Q_2^3 + 2 B Q_2 Q_1^2 - \beta  Q_2^{-3}=0,
\label{eqHH42}
\end{eqnarray}
in which the constants
$A,B,C,\alpha,\beta,\gamma,\Omega_1$ and $\Omega_2$
can only take the four values
(the notation $A:B:C=p:q:r$ stands for $A/p=B/q=C/r=\hbox{arbitrary}$),
\begin{eqnarray}
& & \left\lbrace
\begin{array}{ll}
\displaystyle{
A:B:C=1:2:1,\ \gamma=0,
}
\\
\displaystyle{
A:B:C=1:6:1,\ \gamma=0,\ \Omega_1=\Omega_2,
}
\\
\displaystyle{
A:B:C=1:6:8,\ \alpha=0,\ \Omega_1=4\Omega_2,
}
\\
\displaystyle{
A:B:C=1:12:16,\ \gamma=0,\ \Omega_1=4\Omega_2.
}
\end{array}
\right.
\label{eqHH4NLScond}
\end{eqnarray}
\end{enumerate}

For each of the seven cases so isolated
there exists a second constant of the motion $K$
\cite{Drach1919KdV,BEF1995b,H1984} 
\cite{H1987,BakerThesis,BEF1995b} 
which commutes with the Hamiltonian,
\begin{eqnarray}
{\hskip -00.0truemm}
\hbox{(SK)} : K & = &
K_0^2 + 3 \gamma (3 p_1^2 q_2^{-2} + 4 \alpha q_1 + 2 \omega_2),\
\label{eqHH3SKSecond}
\\ & &
K_0=3 p_1 p_2 + \alpha q_2 (3 q_1^2 + q_2^2) + 3 \omega_2 q_1 q_2,
\nonumber
\\
{\hskip -00.0truemm}
\hbox{(KdV5)} : K & = &
4 \alpha p_2 (q_2 p_1 - q_1 p_2)
+ (4 \omega_2 - \omega_1) (p_2^2 + \omega_2 q_2^2 + \gamma q_2^{-2})
\nonumber \\ & &
+ \alpha^2 q_2^2 (4 q_1^2 + q_2^2)
+ 4 \alpha q_1 (\omega_2 q_2^2 - \gamma q_2^{-2}),
\label{eqHH3K5Second}
\\
{\hskip -00.0truemm}
\hbox{(KK)} : K & = &
(3 p_2^2 + 3 \omega_2 q_2^2 + 3 \gamma q_2^{-2})^2
+ 12 \alpha p_2 q_2^2 (3 q_1 p_2 - q_2 p_1)
\nonumber \\ & &
- 2 \alpha^2 q_2^4 (6 q_1^2 + q_2^2)
+ 12 \alpha q_1 (-\omega_2 q_2^4 + \gamma)
- 12 \omega_2 \gamma,
\label{eqHH3KKSecond}
\\
{\hskip -00.0truemm}
\hbox{quartic} : K & = & \hbox{see }
(\ref{eqHH401:2:1K}), (\ref{eqHH40161}), (\ref{eqHH40168}), (\ref{eqHH401:12:16K}).
\end{eqnarray}
Therefore
one should be able to integrate both in the
Hamilton-Jacobi sense (separation of variables)
and in the Painlev\'e sense (closed-form single-valued general solution).
This invariant $K(q_1,q_2,p_1,p_2)$ is polynomial in the momenta $p_1,p_2$,
with the degrees 2 (KdV5 and 1:2:1 cases) and 4 (the five other cases)
and the difficulty to perform the separation of variables is intimately
related to the degree of $K$ in the momenta.

\section     {Link to soliton equations}
\label{sectionLink}
\indent

In the cubic case
it is possible to build \cite{Fordy1991}
by elimination of $q_2$ between the three equations (\ref{eqHH0})--(\ref{eqHH2}),
a fourth-order ODE for $q_1(t)$
with two nice properties:
\begin{enumerate}
\item
$q_1''''$ is a polynomial in $q_1''',q_1'',q_1',q_1$,
without the ${q_1'''}^2$ term,
\item
this fourth-order ODE is, in each of the three cases,
the traveling wave reduction of a fifth-order soliton equation.
\end{enumerate}
This ODE, namely
\begin{eqnarray}
& &
 q_1'''' + (8 \alpha - 2 \beta) q_1 q_1''
 - 2 (\alpha + \beta) q_1'^2
 - \frac{20}{3} \alpha \beta q_1^3
\nonumber
\\
& &
  +(\omega_1 + 4 \omega_2) q_1''
  + (6 \alpha \omega_1 - 4 \beta \omega_2) q_1^2 + 4 \omega_1 \omega_2 q_1
   + 4 \alpha E
=0,
\label{eqHH3ODE4}
\end{eqnarray}
is independent of the coefficient $\gamma$ of the nonpolynomial term $q_2^{-2}$
and it depends on the constant value $E$ of the Hamiltonian $H$.

This elimination establishes the identification
\cite{Fordy1991}
of the HH3 Hamiltonian system
with the traveling wave reduction $u(x,t)=U(x-ct)$
of the fifth-order conservative
partial differential equation (PDE)
\begin{eqnarray}
& &
u_t + \Big(u_{xxxx} + (8 \alpha - 2 \beta) u u_{xx}
 - 2 (\alpha + \beta) u_x^2
 - \frac{20}{3} \alpha \beta u^3\Big)_x = 0,
\label{eqHHPDE}
\end{eqnarray}
and the three values of $\beta/ \alpha,\omega_1$ and $\omega_2$
for which HH3 passes the Painlev\'e test
are precisely the only values for which the PDE
(\ref{eqHHPDE}) is a soliton equation,
respectively called
the Sawada-Kotera (SK) \cite{SK1974}
\index{Sawada-Kotera equation},
fifth-order KdV (KdV5) \cite{Lax}
\index{fifth-order KdV equation}
and
Kaup-Kupershmidt (KK) \cite{Kaup1980,FG1980a}
\index{Kaup-Kupershmidt equation}
equations.

In each of the four quartic cases
one can similarly establish a link \cite{FK1983,BakerThesis} with
a soliton system made of two coupled PDEs,
most of them appearing in lists established from group theory
\cite{DS1981}.
However,
the elimination of $Q_2$ in a way similar to the cubic case
leads to
\begin{eqnarray}
& &
-Q_1''''
+ 2 \frac{Q_1' Q_1'''}{Q_1}
+\left(1 + 6 \frac{A}{B}\right) \frac{{Q_1''}^2}{Q_1}
-2 \frac{{Q_1'}^2 Q_1''}{Q_1^2}
\nonumber \\ & & \phantom{12345}
+8 \left(6 \frac{A C}{B} - B - C\right) Q_1^2 Q_1''
+4(B - 2 C) Q_1 {Q_1'}^2
+24 C \left(4 \frac{A C}{B} - B\right) Q_1^5
\nonumber \\ & & \phantom{12345}
+\left\lbrack
12 \frac{A}{B} \omega_1 - 4 \omega_2
+\left(1 + 12 \frac{A}{B}\right) \frac{\gamma}{Q_1}
- 4 \left(1+3 \frac{A}{B}\right) \frac{\alpha}{Q_1^4}
\right\rbrack Q_1''
\nonumber \\ & & \phantom{12345}
+ 6 \frac{A}{B} \frac{\alpha^2}{Q_1^7}
+ 20 \frac{\alpha} {Q_1^5} {Q_1'}^2
-12  \frac{A}{B} \frac{\gamma \alpha}{Q_1^4}
+4 \left(3 \frac{A}{B} \omega_1 - \omega_2 \right)
   \left(\gamma - \frac{\alpha}{Q_1^3}\right)
-2 \gamma \frac{{Q_1'}^2}{Q_1^2}
\nonumber \\ & & \phantom{12345}
+ 6 \left(\frac{A}{B} \gamma^2 + 2 B \alpha -8 \frac{A C}{B} \alpha\right)
     \frac{1}{Q_1}
+ \left(6 \frac{A}{B} \omega_1^2 -4 \omega_1 \omega_2 -8 B E\right) Q_1
\nonumber \\ & & \phantom{12345}
+ 48 \frac{A C}{B} \gamma Q_1^2
+ 4 \left(12  \frac{A C}{B} - B - 4 C \right) \omega_1 Q_1^3=0.
\label{eqHH4odeq1}
\end{eqnarray}
This ODE, which depends on $E$ but not on $\beta$,
is equivalent to the Hamilton's equations.
Therefore this would be the most suitable ODE to which to
apply the Painlev\'e test.

In the 1:12:16 case with the constraint $\alpha=0$
this ODE is identical to the autonomous restriction of
\cite[Eq.~(5.9)]{KitaevP2},
an equation linked to the hierarchy of the second Painlev\'e equation,
reproduced as \cite[Eq.~(7.141)]{Cos2000c}.
The Hamiltonian system
equivalent to this ODE is easily integrated
by the method of separation of variables \cite{AF2000},
see section \ref{sectionIntegrationQuartic1:12:16}.
The results to be displayed in next sections
show that, in the four HH4 cases,
the general solution $Q_1^2$ of (\ref{eqHH4odeq1}) is single-valued,
with in addition $Q_1$ single-valued in the 1:6:8 case.

\section     {Strategies to perform the explicit integration}
\label{sectionStrategies}
\indent

In order to find the general solution in closed form
for each of the seven cases
one can think of three strategies.
By decreasing order of elegance
these are the following.
\begin{enumerate}
\item
Take advantage of the knowledge of the second invariant $K$
(integrability in the Liouville sense)
to find a canonical transformation to separating variables,
i.e.~to integrate in the Hamilton-Jacobi sense
and then prove that the Hamilton's equations
written for the separating variables
have a single-valued general solution.
This is the natural strategy,
but is also the most difficult one.
\item
To eliminate one of the two variables, say $q_2(t)$,
between the two Hamilton's equations
and the two constants of the motion
and to identify one of the three resulting ODEs for, say $q_1(t)$,
as a member in a list of ODEs already classified and integrated
by classical authors like
Chazy \cite{ChazyThese}, Bureau \cite{BureauMII}
or Cosgrove \cite{Cos2000a,Cos2000c}.
The main difficulty is that,
since
systems of two coupled second-order ODEs have not yet been classified,
one must eliminate one of the two variables,
which in the quartic case
generates a nonpolynomial ODE such as (\ref{eqHH4odeq1}),
which has not yet been classified.
\item
To establish a birational transformation
between a classified ODE and one of the seven cases
and then carry out the solution.
\end{enumerate}

\section     {Integration of the HH3-KdV5 and HH4-1:2:1 cases}
\label{sectionIntegrationStaeckel}
\indent

When the degree of $K$ is two,
there exists a general method \cite{Staeckel1893}
to find the separating variables
and we just recall its results for completeness.

\subsection{The cubic case $\beta / \alpha=-6$ (KdV5)}

Under the canonical transformation
to parabolic coordinates
\cite{Drach1919KdV,AP1983,Woj1984},
\begin{eqnarray}
& &
(q_1,q_2,p_1,p_2) \to (s_1,s_2,r_1,r_2),
\\
& &
q_1 = -(s_1+s_2+\omega_1-4 \omega_2)/(4 \alpha),\
q_2^2=- s_1 s_2 /(4 \alpha^2),\
\\
& &
p_1=-4 \alpha\frac{s_1 r_1 - s_2 r_2}{s_1-s_2},\
p_2^2=-16 \alpha^2 \frac{s_1 s_2 (r_1-r_2)^2}{(s_1-s_2)^2},
\end{eqnarray}
the Hamiltonian takes the form
\begin{eqnarray}
& &
H=\frac{f(s_1,r_1)-f(s_2,r_2)}{s_1-s_2},\
\\
& &
f(s,r)=-\frac
  {s^2 (s+\omega_1-4 \omega_2)^2 (s-4 \omega_2)-64 \alpha^4 \gamma}{32 \alpha^2 s}
  +8 \alpha^2 r^2 s.
\end{eqnarray}
Therefore the Hamilton-Jacobi equation (\ref{eqHJAutonomous})
allows the introduction of a separating constant $K$
identical to the second
constant of the motion (\ref{eqHH3K5Second}) so that
\begin{eqnarray}
& &
f(s_j,r_j)- E s_j + \frac{K}{2}=0,\ j=1,2.
\label{eqHH3K5Separated}
\end{eqnarray}
The transformed Hamilton's equations
\begin{eqnarray}
& &
s_1'=\pard{H}{r_1}= 16 \alpha^2 \frac{s_1}{s_1-s_2} r_1,\
s_2'=\pard{H}{r_2}= 16 \alpha^2 \frac{s_2}{s_2-s_1} r_2,
\end{eqnarray}
are equivalently written as
\begin{eqnarray}
& &
(s_1-s_2)s_1'=\sqrt{P(s_1)},\
(s_2-s_1)s_2'=\sqrt{P(s_2)},\
\label{eqHyperellipticGenusTwoSystem}
\\
& &
P(s)=s^2 (s+\omega_1-4 \omega_2)^2 (s-4 \omega_2)
       + 32 \alpha^2 E s^2 - 16 \alpha^2 K s -64 \alpha^4 \gamma,
\end{eqnarray}
called a
\textit{hyperelliptic system} of genus two.
 \index{hyperelliptic!system}
The variables $q_1$ and $q_2^2$ are meromorphic and the Hamiltonian system has
the Painlev\'e property.

\subsection{The quartic case 1:2:1}

\begin{eqnarray}
H & = &
 \frac{1}{2}(p_1^2+p_2^2)
+\frac{1}{2}(\omega_1 q_1^2+\omega_2 q_2^2)
+\frac{1}{2} (q_1^4+ 2 q_1^2 q_2^2 + q_2^4)
\nonumber
\\
& &
 +\frac{1}{2}\left(\frac{\alpha}{q_1^2}+\frac{\beta}{q_2^2}\right)=E,
\label{eqHH401:2:1E}
\\
K & = &
\left(q_2 p_1 - q_1 p_2 \right)^2
+ q_2^2 \frac{\alpha}{q_1^2} + q_1^2 \frac{\beta}{q_2^2}
\nonumber
\\
& &
-\frac{\omega_1-\omega_2}{2}
\left(p_1^2-p_2^2+q_1^4-q_2^4+\omega_1 q_1^2 - \omega_2 q_2^2
+ \frac{\alpha}{q_1^2} - \frac{\beta}{q_2^2}
 \right).
\label{eqHH401:2:1K}
\end{eqnarray}
Quite similarly
the canonical transformation to elliptic coordinates \cite{Woj1985},
\begin{eqnarray}
& &
\left\lbrace
\begin{array}{ll}
\displaystyle{
q_j^2=(-1)^j \frac{(s_1+\omega_j)(s_2+\omega_j)}{\omega_1-\omega_2},\ j=1,2,
}
\\
\displaystyle{
p_j=2 q_j \frac{\omega_{3-j} (r_2-r_1) -s_1 r_1 + s_2 r_2}{s_1-s_2},\ j=1,2
},
\end{array}
\right.
\end{eqnarray}
maps the Hamilton's equations to the hyperelliptic system
(\ref{eqHyperellipticGenusTwoSystem}) with
\begin{eqnarray}
P(s)&=&
s(s+\omega_1)^2(s+\omega_2)^2
-\alpha (s+\omega_2)^2-\beta (s+\omega_1)^2
\nonumber
\\
& &
-(s+\omega_1)(s+\omega_2)\left[E (2s+\omega_1+\omega_2)-K\right].
\end{eqnarray}

We remark that
the variable $x=q_1^2+q_2^2$ obeys the fourth-order ODE
\begin{eqnarray}
& &
x'''' + (20 x +4 \omega_1+4 \omega_2) x'' + 10 {x'}^2 + 40 x^3
\nonumber
\\
& &
+ 8 (\omega_1+\omega_2)(3 x^2 -E)
+ (16 \omega_1 \omega_2 - E) x
- 8 (\alpha+\beta+K)=0,
\end{eqnarray}
which, up to some translation,
is identical to the ODE (\ref{eqHH3ODE4}) in the KdV5 case.

\section     {Integration of the cubic cases SK and KK}
\label{sectionIntegrationCubicSKandKK}
\indent
\begin{eqnarray}
H_{\rm SK}
& =&
 \frac{1}{2} (P_1^2 + P_2^2)  + \frac{\Omega_1}{2} (Q_1^2 + Q_2^2)
+ \frac{1}{2} Q_1 Q_2^2 + \frac{1}{6} Q_1^3 + \frac{\lambda^2}{8} Q_2^{-2},\
\label{eqHH3SK}
\\
H_{\rm KK}
& =&
 \frac{1}{2} (p_1^2 + p_2^2) + \frac{\omega_2}{2} (16 q_1^2 + q_2^2)
    + \frac{1}{4} q_1 q_2^2 + \frac{4}{3} q_1^3 + \frac{\lambda^2}{2} q_2^{-2}.
\label{eqHH3KK}
\end{eqnarray}
These two cases are equivalent under a birational canonical transformation
\cite{BW1994,SEL},
which exchanges the two sets $(H,K,\Omega_1,\lambda^2)_{\rm SK}$
and $(H,K,\omega_2,\lambda^2)_{\rm KK}$.
The two Hamilton-Jacobi equations are simultaneously separated as follows
\cite{RGC,VMC2002a}.
\begin{enumerate}
\item
One introduces the canonical transformation to Cartesian coordinates
\begin{eqnarray}
& & \left\lbrace
\begin{array}{ll}
\displaystyle{
\QA=Q_1 + \Omega_1 + Q_2,\ \PA=(P_1+P_2)/2,\
}
\\
\displaystyle{
\QB=Q_1 + \Omega_1 - Q_2,\ \PB=(P_1-P_2)/2,
}
\end{array}
\right.
\label{eqTCSKSK}
\end{eqnarray}
which trivially separates $H_{\rm SK}$ for $\lambda=0$,
\begin{eqnarray}
& & \lambda=0:\ H_{\rm SK} =\PA^2 + \PB^2 + \frac{1}{12}(\QA^3 + \QB^3)
-4 \Omega_1^2 (\QA + \QB).
\end{eqnarray}
\item
One then applies to $H_{\rm KK}$
two canonical transformations,
firstly the transformation $(q_j,p_j) \to (Q_j,P_j)$ 
taken for $\lambda=0$ and
secondly the rotation (\ref{eqTCSKSK}),
which results in
\begin{eqnarray}
{\hskip -15.0truemm}
& &
q_1=-6 \left(\frac{\PA - \PB}{\QA - \QB} \right)^2 -\frac{\QA + \QB}{2},\
q_2^2=24 \frac{f(\QA,\PA)-f(\QB,\PB)}{\QA - \QB},
\\
{\hskip -15.0truemm}
& &
p_1=-4 \QA \frac{\PA - \PB}{\QA - \QB} - 2 \frac{\QA \PB - \QB \PA}{\QA - \QB},\
p_2= \QB \frac{\PA - \PB}{\QA - \QB},
\\
{\hskip -15.0truemm}
& &
H_{\rm KK}=f(\QA,\PA)+f(\QB,\PB)
 + \frac{\lambda^2}{24} \frac{\QA-\QB}{f(\QA,\PA)-f(\QB,\PB)},
\label{eqHH3HamKKnew}
\\
{\hskip -15.0truemm}
& &
f(q,p)=p^2+\frac{1}{12}q^3-4 \omega_2^2 q.
\end{eqnarray}
Therefore both Hamilton-Jacobi equations are separated \cite{VMC2002a},
viz.
\begin{eqnarray}
& &
(f(\Qj,\Pj) - E/2)^2 + (\lambda^2/24) \Qj + K=0,\ j=1,2,
\end{eqnarray}
with $K$ the second integral of the motion
(\ref{eqHH3SKSecond}) or (\ref{eqHH3KKSecond}).
In the particular case $\lambda=0$,
the Hamiltonians themselves are separated \cite{RGC}.
\item
Finally the Hamilton's equations in the variables $(\QA,\QB,\PA,\PB)$
are identified \cite{VMC2002a} to a hyperelliptic system
of the canonical form
(\ref{eqHyperellipticGenusTwoSystem}),
\begin{eqnarray}
& & \left\lbrace
\begin{array}{ll}
\displaystyle{
\QA=s_1^2- \frac{3 K}{\lambda^2},\
\QB=s_2^2- \frac{3 K}{\lambda^2},\
\PA=\frac{r_1}{2 s_1},\ 
\PB=\frac{r_2}{2 s_2},\
}
\\
\displaystyle{
P(s)=-\frac{1}{3}\left(s^2-3 \frac{K}{\lambda^2}\right)^3
    + \Omega_1^2 \left(s^2-3 \frac{K}{\lambda^2}\right)
 + \frac{\lambda}{\sqrt{3}} s + 2 E,
}
\end{array}
\right.
\label{eqHH3SKKKHyperCurve}
\end{eqnarray}
thus providing the meromorphic general solution
\begin{eqnarray}
& & \left\lbrace
\begin{array}{ll}
\displaystyle{
q_1=- \frac{s_1^2+s_2^2}{2}-\frac{3}{2}\left(\frac{s_1'+s_2'}{s_1+s_2}\right)^2
+ \frac{3 K}{\lambda^2},\
}
\\
\displaystyle{
q_2^{-2}=\frac{s_1+s_2}{2 \sqrt{3} \lambda},
}
\\
\displaystyle{
Q_1= \sqrt{3}(s_1'+s_2')
  +s_1^2+s_2^2+s_1 s_2  - \frac{3 K}{\lambda^2},\
}
\\
\displaystyle{
Q_2^2=-2 \sqrt{3}(s_1+s_2)(s_1 s_1'+s_2 s_2')
+2(s_1+s_2)^2\left(s_1^2+s_2^2-\frac{9 K}{2 \lambda^2}\right).
}
\end{array}
\right.
\end{eqnarray}
\end{enumerate}
{\vskip -2.0truemm}
\textit{Remark}.
Cosgrove \cite{Cos2000a} was the first to obtain the
above hyperelliptic expressions for $q_1$ and $Q_1$,
by a direct integration of the fourth-order ODE (\ref{eqHH3ODE4})
in the KK and SK cases.
They are respectively denoted F-III and F-IV in his classification
and $\lambda^2$ is a first integral of the ODE.
Therefore setting $\lambda=0$ would prevent finding its general solution.

\section     {Integration of the quartic 1:6:1 and 1:6:8 cases}
\label{sectionIntegrationQuartic1:6:1and1:6:8}
\indent
\begin{eqnarray}
{\hskip -15.0 truemm}
& &
1:6:1
\left\lbrace
\begin{array}{ll}
\displaystyle{
H =
 \frac{1}{2}(P_1^2+P_2^2)
+\frac{\omega_1}{2}(Q_1^2+Q_2^2)
-\frac{1}{32} (Q_1^4+ 6 Q_1^2 Q_2^2 + Q_2^4)
}
\\
\displaystyle{
\phantom{1234}
 -\frac{1}{2}\left(\frac{\kappa_1^2}{Q_1^2}+\frac{\kappa_2^2}{Q_2^2}\right)
=E,
}
\\
\displaystyle{
K =
\left(
P_1 P_2 + Q_1 Q_2 \left(-\frac{Q_1^2+Q_2^2}{8}+\omega_1 \right)
\right)^2
}
\\
\displaystyle{
\phantom{1234}
- P_2^2 \frac{\kappa_1^2}{Q_1^2}
- P_1^2 \frac{\kappa_2^2}{Q_2^2}
+\frac{1}{4}\left(\kappa_1^2 Q_2^2 + \kappa_2^2 Q_1^2 \right)
+\frac{\kappa_1^2 \kappa_2^2}{Q_1^2 Q_2^2}
}
\end{array}
\right.
 \label{eqHH40161}
\end{eqnarray}
and
\begin{eqnarray}
{\hskip -15.0 truemm}
& &
1:6:8
\left\lbrace
\begin{array}{ll}
\displaystyle{
H =
 \frac{1}{2}(p_1^2+p_2^2)
+\frac{\omega_2}{2}(4 q_1^2+q_2^2)
-\frac{1}{16} (8 q_1^4+ 6 q_1^2 q_2^2 + q_2^4)
}
\\
\displaystyle{
\phantom{1234}
+ \gamma q_1 +\frac{\beta}{2 q_2^2}
=E,
}
\\
\displaystyle{
K =
\left(
p_2^2-\frac{q_2^2}{16}(2 q_2^2+4 q_1^2+\omega_2)
     +\frac{\beta}{q_2^2}
\right)^2
-\frac{1}{4}q_2^2(q_2 p_1 - 2 q_1 p_2)^2
}
\\
\displaystyle{
\phantom{1234}
+\gamma
\left(
-2 \gamma q_2^2
-4 q_2 p_1 p_2
+\frac{1}{2} q_1 q_2^4
+ q_1^3 q_2^2
+4 q_1 p_2^2
-4 \omega_2 q_1 q_2^2
+ 4 q_1 \frac{\beta}{q_2^2}
\right).
}
\end{array}
\right.
\label{eqHH40168}
\end{eqnarray}
The situation is similar to that for the cubic SK and KK cases.
There is a canonical transformation \cite{BakerThesis}
between the 1:6:1 and 1:6:8 cases which maps the constants as follows
\begin{eqnarray}
H_{1:6:8}=H_{1:6:1},\
K_{1:6:8}=K_{1:6:1},\
\omega_2=\omega_1,\
\gamma=\frac{\kappa_1+\kappa_2}{2},\
\beta=- (\kappa_1-\kappa_2)^2.
\label{eqCT161to168constants}
\end{eqnarray}
However, the separating variables have only been found for $\beta \gamma=0$.
In the case $\beta=\gamma=0$ \cite{BSV1982,AP1983}
the canonical transformation,
\begin{eqnarray}
& &
\left\lbrace
\begin{array}{ll}
\displaystyle{
\QA=\frac{1}{2} (Q_1+Q_2)^2,\
\QB=\frac{1}{2} (Q_1-Q_2)^2,\
}
\\
\displaystyle{
\PA= \frac{P_1+P_2}{2(Q_1+Q_2)},\
\PB= \frac{P_1-P_2}{2(Q_1-Q_2)},\
}
\end{array}
\right.
\label{eqCT168particular}
\end{eqnarray}
separates the Hamiltonian $H_{1:6:1}$
\begin{eqnarray}
{\hskip -10.0 truemm}
& & \kappa_1=\kappa_2=0:\
H_{1:6:1}
= f(\QA,\PA) + f(\QB,\PB),\
f(q,p)=2 q p^2 - \frac{1}{16} q^2 + \frac{\omega_1}{2} q
\label{eq161f}
\end{eqnarray}
and leads to elliptic functions for $Q_1,Q_2,q_1,q_2^2$.
In the generic case
the best achievement to date for the separating variables \cite{RRG}
is to proceed as in the cubic SK-KK case.
After applying two canonical transformations,
firstly the transformation $(q_j,p_j) \to (Q_j,P_j)$
taken for $\beta=\gamma=\kappa_1=\kappa_2=0$
and secondly the transformation (\ref{eqCT168particular}),
the Hamilton-Jacobi equation $H_{1:6:8}-E=0$ becomes
\begin{eqnarray}
& &
\left\lbrace
\begin{array}{ll}
\displaystyle{
g(\QA,\PA)-g(\QB,\PB)
 - \gamma \sqrt{\QA \QB} \frac{\PA-\PB}{\QA-\QB}
   \left(f(\QA,\PA)-f(\QB,\PB)\right)
=0,
}
\\
\displaystyle{
g(q,p)=\frac{1}{4} f(q,p)^2 - E f(q,p) + \frac{\beta}{8} q,\
f(q,p)=2 q p^2 - \frac{1}{16} q^2 + \frac{\omega_1}{2} q,
}
\end{array}
\right.
\end{eqnarray}
i.e., it separates only for $\gamma=0$.
The Hamilton's equations in the variables $(\QA,\QB,\PA,\PB)$
can then be identified \cite{V2003} to a hyperelliptic system
of the canonical form
(\ref{eqHyperellipticGenusTwoSystem})
\begin{eqnarray}
{\hskip -15.0 truemm}
& & \gamma=0:\ 
{\hskip -3.0truemm}
\left\lbrace
\begin{array}{ll}
\displaystyle{
\QA=s_1^2- \frac{K}{2 \kappa_1^2},\
\QB=s_2^2- \frac{K}{2 \kappa_1^2}.
}
\\
\displaystyle{
P(s)=\frac{1}{2}\left(s^2- \frac{K}{2 \kappa_1^2}\right)^3
  -4 \omega_1^2 \left(s^2- \frac{K}{2 \kappa_1^2}\right)^2
 +\left(4 E + 2 \sqrt{2} \kappa_1 s\right)
                \left(s^2- \frac{K}{2 \kappa_1^2}\right)
}
\end{array}
\right.
\label{eq1:6:1gamma0HyperEll}
\end{eqnarray}
and thus provides the meromorphic general solution
\begin{eqnarray}
{\hskip -8.0truemm}
& & \gamma=0:\
q_1^2= -\frac{s_1^2+s_2^2}{2}+\left(\frac{s_1'+s_2'}{s_1+s_2}\right)^2
- \frac{2 \sqrt{2} \kappa_1}{s_1+s_2}+ \frac{K}{2 \kappa_1^2} + 4 \omega_1,\
q_2^{2}=\frac{4 \sqrt{2} \kappa_1}{s_1+s_2}.
\label{eq1:6:8gamma0}
\end{eqnarray}
In the generic case $\beta \gamma \not=0$
the second strategy (see section \ref{sectionStrategies}) cannot be used
since the ODE (\ref{eqHH4odeq1}) belongs to a class not yet
investigated for the Painlev\'e property.
Fortunately the third strategy succeeds in performing the integration
and one can establish a birational transformation between
the ODE (\ref{eqHH4odeq1}) and the autonomous F-VI equation (a-FVI)
in the classification of Cosgrove \cite{Cos2000a},
viz.
\begin{eqnarray}
& &
\hbox{a-F-VI}:\
y''''=18 y y'' + 9 {y'}^2 - 24 y^3
+ \alpha_{\rm VI} y^2 + \frac{\alpha_{\rm VI}^2}{9} y
+ \kappa t + \beta_{\rm VI},\
\kappa=0,
\label{eqCosgroveFVI}
\end{eqnarray}
an ODE the general solution of which is meromorphic and
expressed with genus two hyperelliptic functions \cite[Eq.~(7.26)]{Cos2000a}.
The principle, explained in \cite{MV2003,VMC2004a},
is to remark that the 1:6:8 Hamilton's equations
and the a-F-VI ODE
are the traveling wave reduction
of two soliton systems linked by a B\"acklund transformation (BT).
These are, respectively,
the coupled KdV system denoted \cKdV${}_1$
\cite{BEF1995b,BakerThesis}, viz.
\begin{eqnarray}
{\hskip -10.0 truemm}
& &
\left\lbrace
\begin{array}{ll}
\displaystyle{
f_\tau+ \left(f_{xx}+\frac{3}{2} f f_{x}-\frac{1}{2}f^3+3 f g\right)_x=0,
}
\\
\displaystyle{
- 2 g_\tau+ g_{xxx}+ 6 g g_x+3 f g_{xx}+6 gf_{xx}+9 f_x g_x-3 f^2 g_x
}
\\
\displaystyle{
\phantom{xxxxx}+\frac{3}{2} f_{xxxx} +\frac{3}{2} f f_{xxx}+9 f_x f_{xx}
               -3 f^2 f_{xx}-3 f f_x^2=0,
}
\end{array}
\right.
\label{eq:cKdV1}
\end{eqnarray}
and another system of the \cKdV\ type, denoted bi-SH system
\cite{DS1981,SH1982,JM1983,DS1984},
\begin{eqnarray}
& &
\left\lbrace
\begin{array}{ll}
\displaystyle{
- 2 u_\tau+ \left(u_{xx} + u^2 + 6 v\right)_x = 0,
}
\\
\displaystyle{
v_\tau+ v_{xxx} + u v_x = 0.
}
\end{array}
\right.
\label{eq:systemHSII}
\end{eqnarray}
This BT is defined by the Miura transformation
\begin{eqnarray}
{\hskip -10.0 truemm}
& &
\left\lbrace
\begin{array}{ll}
\displaystyle{
u=\frac{3}{2} \left(2 g -f_x-f^2\right),
}
\\
\displaystyle{
}
\\
\displaystyle{
v=\frac{3}{4}
\left(
 2 f_{xxx}
+4 f f_{xx}
+8 g f_x
+4 f g_x
+3 f_x^2
-2 f^2 f_x
-  f^4
+4 g f^2\right).
}
\end{array}
\right.
\end{eqnarray}
Under the reduction $x - c \tau =t$
the B\"acklund transformation between the two PDE systems
becomes a birational transformation between 1:6:8
and the a-F-VI equation,
see details in \cite{CMVNanjing2004}.
The result is a meromorphic general solution
$Q_1^2,Q_2^2,q_1,q_2^2$,
rationally expressed as
\begin{eqnarray}
{\hskip -10.0 truemm}
& &
\left\lbrace
\begin{array}{ll}
\displaystyle{
q_1=\frac{W'}{2 W}
+ \frac{\gamma}{W}
\left[9 j -3 \left(y + \frac{4}{9}\omega_2\right) (h+E)
      - \frac{9}{4} \gamma^2\right],
}
\\
\displaystyle{
q_2^2=-16 \left(y - \frac{5}{9} \omega_2\right)
}
\\
\displaystyle{
\phantom{123}
+\frac{1}{W}
\Big[
 12 \left(y' + \frac{\gamma}{2}\right)^2
 -48 y^3 - 16 \omega_2 y^2
 +\left(24 E +\frac{128}{9}\omega_2^2\right) y
 + \frac{1280}{243} \omega_2^3
\Big.
} \\ \displaystyle{\phantom{12345678}
\Big.
- \frac{40}{3} \omega_2 E + \frac{3}{4} \beta
-24 \gamma \left(y - \frac{5}{9} \omega_2\right) h'
-144 \gamma^2 \left(y - \frac{5}{9} \omega_2\right)^2
\Big],
}
\\
\displaystyle{
W= (h+E)^2 -9 \gamma^2 \left(y - \frac{5}{9} \omega_2\right),
}
\\
\displaystyle{
\alpha_{\rm VI}=4 \omega_2,\
\beta_{\rm VI}=\frac{3}{4} \gamma^2 + 2 \omega_2 E
               - \frac{3}{16} \beta-\frac{512}{243} \omega_2^3,
}
\end{array}
\right.
\label{eqFVI_to_HH4168}
\end{eqnarray}
in which $h$ and $j$ are convenient auxiliary variables
\cite[Eqs.~(7.4)--(7.5)]{Cos2000a},
\begin{eqnarray}
{\hskip -10.0 truemm}
& &
\left\lbrace
\begin{array}{ll}
\displaystyle{
h=y''-6 y^2 - \frac{4}{3} \omega_2 y + \frac{16}{27} \omega_2^2,
}
\\
\displaystyle{
j=\left(y-\frac{2}{9} \omega_2 \right)y''-\frac{1}{2} {y'}^2 -4 y^3
} \\ \displaystyle{\phantom{123}
+\frac{1}{6}
 \left(4 \omega_2 y^2 + \frac{16}{9} \omega_2^2 y
       -\frac{512}{243} \omega_2^3 + 2 \omega_2 E + \frac{3}{4} \gamma^2
       - \frac{3}{16} \beta\right).
}
\end{array}
\right.
\end{eqnarray}
This shows that $q_1^2$ in (\ref{eq1:6:8gamma0})
is the square of the single-valued expression
\begin{eqnarray}
& & \gamma=0:\
q_1=\frac{h'}{h+E},\ q_2^2=\frac{E}{s_1+s_2}.
\end{eqnarray}
Contrary to previous cases
the coefficients of the hyperelliptic curve \cite[Eq.~(7.23)]{Cos2000a}
depend algebraically \cite{CMVGallipoli2004}
on the parameters $\beta,\gamma,\kappa_1,\kappa_2$ of the Hamiltonians,
and this could explain the difficulty
to separate the variables in the Hamilton-Jacobi equation.
Note that,
in the particular case $\beta \gamma=0$, i.e.~$\kappa_1^2=\kappa_2^2$,
these coefficients become rational as in (\ref{eq1:6:1gamma0HyperEll}).

\section     {Integration of the quartic 1:12:16 case}
\label{sectionIntegrationQuartic1:12:16}
\indent
\begin{eqnarray}
& &
\left\lbrace
\begin{array}{ll}
\displaystyle{
H =
\frac{1}{2}(p_1^2+p_2^2) + \frac{\omega_1}{8} (4 q_1^2+ q_2^2)
 - \frac{1}{32} (16 q_1^4+ 12 q_1^2 q_2^2 + q_2^4)
}
\\
\displaystyle{
\phantom{1234}
 +\frac{1}{2}\left(\frac{\alpha}{q_1^2}+\frac{\beta}{q_2^2}\right)
=E,
}
\\
\displaystyle{
K =
\left(8 (q_2 p_1 - q_1 p_2) p_2 - q_1 q_2^4 - 2 q_1^3 q_2^2
 + 2 \omega_1 q_1 q_2^2 - 8 q_1 \frac{\beta}{q_2^2} \right)^2
}
\\
\displaystyle{
\phantom{1234}
+\frac{32 \alpha}{5}
\left(q_2^4 + 10 \frac{q_2^2 p_2^2}{q_1^2}\right).
}
\end{array}
\right.
\label{eqHH401:12:16E}
\label{eqHH401:12:16K}
\end{eqnarray}
Up to now separating variables are only known in the case $\alpha \beta=0$.
The case $\alpha=0$ belongs to the St\"ackel class
(two invariants quadratic in $p_1,p_2$).
Under the canonical transformation to parabolic coordinates,
\begin{eqnarray}
& &
q_1=s_1+s_2,\ q_2^2=-4 s_1 s_2,\
p_1=\frac{s_1 r_1 - s_2 r_2}{s_1-s_2},\
p_2=q_2 \frac{r_1 - r_2}{2(s_1-s_2)},\
\end{eqnarray}
the Hamilton-Jacobi equation is separated
and $(s_1,s_2)$ obey the system (\ref{eqHyperellipticGenusTwoSystem}) with
\begin{eqnarray}
& &
\alpha=0:\
P(s)=s^6 - \omega_1 s^3 + 2 E s^2 + \frac{K}{20} s - \frac{\beta}{4}.
\label{eq1:12:16alpha0}
\end{eqnarray}
In the case $\beta=0$
the Hamilton-Jacobi equation is separated \cite{V2003} by two successive
canonical transformations
which yield a similar hyperelliptic curve
\begin{eqnarray}
\beta=0:\
P(s)=s^6 - \omega_1 s^3 + 2 E s^2 + \frac{K}{20} s - \alpha.
\label{eq1:12:16beta0}
\end{eqnarray}
In the generic case $\alpha \beta \not=0$,
following the third strategy,
one has found \cite{BakerThesis,V2003,MV2003}
a path the segments of which are either traveling wave reductions
or B\"acklund transformations,
linking the 1:12:16 Hamiltonian
to a hyperelliptic system of the canonical form
(\ref{eqHyperellipticGenusTwoSystem})
with the hyperelliptic curve (\ref{eqHH3SKKKHyperCurve}),
which separates both the cubic SK and KK cases.
However, since the curve (\ref{eqHH3SKKKHyperCurve})
contains neither (\ref{eq1:12:16alpha0}) nor (\ref{eq1:12:16beta0}),
this path is certainly not the optimal one to reach the separating variables.
Nevertheless
this proves the singlevaluedness of the general solution $q_1^2,q_2^2$,
the explicit expression of which
results from the product of the six pieces
\cite{BakerThesis,V2003,VMC2004a}
\begin{eqnarray}
{\hskip -10.0 truemm}
& &
\left\lbrace
\begin{array}{ll}
\displaystyle{
q_1^2=\frac{1}{5} ( 2 R - 6 S + \omega_1),\
q_2^2=\frac{4}{5} (-3 R - 4 S + \omega_1),
}
\\
\displaystyle{
R= W_1' - W_1^2,\
S=-W_2'-\frac{1}{2} W_2^2,
}
\\
\displaystyle{
 W_1=\frac{Q_1}{2}+\frac{Q_2'}{Q_2}-\frac{K_3}{Q_2^2},\
 W_2=\frac{Q_1}{2}-\frac{Q_2'}{Q_2}+\frac{K_3}{Q_2^2},\
K_3=\sqrt{-\alpha}+\frac{1}{2}\sqrt{-\beta},
}
\\
\displaystyle{
Q_1=F,\
Q_2^2=\frac{2}{5}\left(F'-2 F^2 -G + \omega_1\right),
}
\\
\displaystyle{
F,G=\hbox{see below, Eqs.~}(\ref{eqbira_from_cKdVa_to_biSK}),
}
\\
\displaystyle{
U=-3 \left(y - \frac{\omega_1}{30}\right),\
V=-6 y'' +18 y^2 - \frac{9}{5} \omega_1 y +\frac{1}{10} \omega_1^2-\frac{3}{5} E,
}
\end{array}
\right.
\label{eq1:12:16_Six_pieces}
\end{eqnarray}
where $y$ obeys the F-IV ODE \cite{Cos2000a},
integrated with genus two hyperelliptic functions.

In the fifth line of (\ref{eq1:12:16_Six_pieces})
the expressions result from the inversion of the
reduction $(u,v,f,g)(x,\tau)=(U,V,F,G)(x+\omega_1 \tau)$
of the Miura transformation
\begin{eqnarray}
& &
u=\frac{3}{10} \left(3 f_x - f^2 + 2 g \right),\
v=\frac{9}{10}\left(f_{xxx} + g_{xx} + f_x g - f g_x -f f_{xx} + g^2\right),
\end{eqnarray}
i.e.,
\begin{eqnarray}
{\hskip -10.0 truemm}
& &
\left\lbrace
\begin{array}{ll}
\displaystyle{
F=-\frac{W'}{2 W} + \Kaone X_2,
}
\\
\displaystyle{
G=-F^2 - X_1 X_2 + \Kaone \frac{54 U'}{X_1}
 - 54 \Kaone  \left(U + \frac{3 \omega_1}{20}\right) \frac{W'}{W X_1}
 + \frac{2}{3}\left(U + \frac{9 \omega_1}{10}\right),
}
\\
\displaystyle{
W=X_1^2 +108 \Kaone^2 \left(U + \frac{3 \omega_1}{20}\right),\
}
\\
\displaystyle{
X_1=V + 2 U^2 - 3 \omega_1 U + \frac{9}{50} \omega_1^2 - \frac{27}{5} E,
}
\\
\displaystyle{
X_2=
 9 \left(
-4 {U'}^2
+ \frac{8}{3} U V
- \frac{8}{25} \omega_1 U^2
+ \frac{2}{5} \omega_1 V
+ \frac{48}{5} E U
\right.
}
\\
\displaystyle{
\left.
\phantom{12345xxxxxxxxxx}
- \frac{42}{25} \omega_1^2 U
+ \frac{9}{8} (4 \alpha + \beta)
- \frac{9}{2} \Kaone^2
+ \frac{36}{25} \omega_1 E
- \frac{27}{125} \omega_1^3
\right).
}
\\
\displaystyle{
\Kaone=\sqrt{-\alpha}-\frac{1}{2}\sqrt{-\beta}.
}
\end{array}
\right.
\label{eqbira_from_cKdVa_to_biSK}
\end{eqnarray}

\section{Conclusion and open problems}
\label{sectionOpen_problems}

The present results are twofold.
\begin{enumerate}
\item
In the seven cases the general solution is
reducible to a canonical hyperelliptic system with genus two
and therefore meromorphic.
\item
Since each of the seven cases can be mapped to a fourth-order ODE
which is complete in the Painlev\'e sense,
it is impossible to add any term to the Hamiltonian without destroying
the Painlev\'e property.
The seven H\'enon-Heiles Hamiltonians are \textit{complete}.
\end{enumerate}

The main open problems are to find the separating variables in three of the
generic quartic cases.

\section*{Acknowledgments}

The authors acknowledge the financial support of
the Tournesol grant no.~T2003.09
between Belgium and France.
Our thanks also go to the referee who helped us
to improve the text.


\label{lastpage}
\vfill \eject
\end{document}